\documentclass[aps,prd,preprintnumbers,nofootinbib,preprint]{revtex4}
\usepackage{bm}
\usepackage{latexsym}
\usepackage{dcolumn}
\usepackage{amsmath,amsfonts,amssymb}
\usepackage{graphicx,epsfig}
\usepackage{amsthm}

\def\bc {\begin{center}}
\def\ec {\end{center}}
\def\bfg {\begin{figure}}
\def\efg {\end{figure}}
\def\bi {\begin{itemize}}
\def\ei {\end{itemize}}

\newcommand{\be}{\begin{equation}}
\newcommand{\bea}{\begin{eqnarray}}
\newcommand{\eea}{\end{eqnarray}}
\newcommand{\ba}{\begin{array}}
\newcommand{\ea}{\end{array}}
\newcommand{\ee}{\end{equation}}
\newcommand{\bes}{\begin{equation*}}
\newcommand{\beas}{\begin{eqnarray*}}
\newcommand{\eeas}{\end{eqnarray*}}
\newcommand{\bas}{\begin{array*}}
\newcommand{\eas}{\end{array*}}
\newcommand{\ees}{\end{equation*}}
\newcommand{\nn}{\nonumber}
\begin{document}

\title{Searching for $AdS_3$ waves and Asymptotically Lifshitz black holes in $R^3$-NMG}

\author{Giorgos G. Anastasiou $^{1,2}$} \email[email: ]{giorgos.g.anastasiou@gmail.com}
\author{M.R. Setare $^{3}$} \email[email: ]{rezakord@ipm.ir}
\author{Elias C. Vagenas $^{4}$} \email[email: ]{evagenas@gmail.com}


\affiliation{$^1$~Departamento de Ciencias F$\acute{\imath}$sicas, Universidad Andres Bello, Av. Rep$\acute{u}$blica 252, 
Santiago, Chile}

\affiliation{$^2$~Centro de Estudios Cient$\acute{\imath}$ficos (CECs), Casilla 1469, Valdivia, Chile }

\affiliation{$^3$~Department of Science, Campus of Bijar, University of Kurdistan,
Bijar, Iran }

\affiliation{$^4$~Theoretical Physics Group, Department of Physics, Kuwait University, 
P.O. Box 5969, Safat 13060, Kuwait}

\begin{abstract}

\par\noindent
In this paper we  consider the structure of the $AdS_3$ vacua in $R^3$ expansion of the New Massive Gravity ($R^3$-NMG).
We obtain the degeneracies of the $AdS_3$ vacua at several points of the parametric space.
Additionally, following a specific analysis we show that $AdS_3$ wave solutions are present.
Using these wave solutions, we single out two special points of the parametric space for which logarithmic terms appear in the solutions.
The first one is a point at which the effective mass of the wave profile which is interpreted as a scalar mode, completely
saturates the Breitenlohner-Freedman bound of the $AdS_3$ space in which the wave is propagating.
The second special point is a point at which the central charge of the theory vanishes.
Furthermore, we investigate the possibility of asymptotically Lifshitz black solutions to be present in the three-dimensional
$R^3$-NMG. We derive analytically the Lifshitz vacua considering specific relations between the mass parameters of $R^3$-NMG.
A certain polynomial equation arises at the first special point where solutions with logarithmic fall-off in the $AdS_3$ space appear.
Solving this polynomial equation, we obtain the values of the dynamical exponent $z$ which correspond
to possible asymptotically Lifshitz black hole solutions. However, it is shown that asymptotically Lifshitz black solutions
do not exist in the three-dimensional $R^3$-NMG for a specific ansatz of the black hole metric.
\end{abstract}
\maketitle
%
%
%
%
\section{Introduction}
%
%
%
In recent years, some interesting models of massive gravity in
three dimensions have been introduced. Among these models, a well known one is that of the
new massive gravity (NMG) \cite{Bergshoeff:2009hq}. This model is equivalent to the three-dimensional
Fierz-Pauli action for a massive spin-2 field at the linearized level. In addition, NMG preserves parity symmetry which is not the case for
the topological massive gravity (TMG) \cite{Deser:1981wh}.
It has been shown that NMG admits BTZ and warped $AdS_3$ black holes solutions \cite{Bergshoeff:2009aq,Clement:2009gq}.
Moreover, Ayon-Beato et al. have shown that Lifshitz metrics with generic values of
the dynamical exponent $z$ are also exact solutions of NMG \cite{AyonBeato:2009nh}.

It is possible to extend NMG to higher curvature theories.
One of these extension of NMG was presented by Sinha \cite{Sinha:2010ai} in which  $R^3$ terms were added to the action of the theory.
Another extension of NMG to a higher curvature theory is that described by the Born-Infeld (BI)  action \cite{Gullu:2010pc}.
Additionally, Nam, Park, and Yi have found BTZ and warped $AdS_3$ black hole solution in the context of $R^3$-NMG as well as in the BI extension of NMG \cite{Nam:2010dd}.

The remainder of the paper goes as follows. In Section II, we briefly present the action and the equations of motions of $R^3$-NMG.
In Section III, we  describe the structure of the $AdS_3$ vacua in $R^3$- NMG, obtain the degeneracies of the $AdS_3$ vacua and find the point in the parametric
space where the central charge is zero. In Section IV, we derive the $AdS_3$ wave solutions and obtain the two special points  of the parametric space
at which wave solutions with logarithmic terms appear. In Section V, we derive the asymptotically Lifshitz vacua in $R^3$-NMG, study the degeneracy of these vacua and
obtain a polynomial equation which corresponds to the first special point. Solving this polynomial equation, we  get the values of the dynamical exponent $z$
which correspond to possible  asymptotically Lifshitz black hole solutions. Furthermore, we show that for a specific ansatz for the black hole metric there are not
any asymptotically Lifshitz black solutions in the three-dimensional $R^3$-NMG.
Finally, in Section VI, we briefly present and discuss the results of the paper.
%
%
%
%
\section{$R^3$ extension of New Massive Gravity}
%
%
%
\par\noindent
In this section we concisely present the model of $R^3$ extension of New Massive Gravity ($R^3$-NMG).
The action of $R^3$-NMG is  given as \cite{Sinha:2010ai,Nam:2010dd}\\
\be
S=\frac{\eta}{2\kappa^2}{\int d^{3}x\sqrt{-g}\left[\sigma
R-2\Lambda+\frac{1}{m^2}K+\frac{\xi}{12\mu^4}K'\right]}
\,,\label{action1}
\ee
\par\noindent
where $m$, $\mu$ are mass parameters of the theory, $2\kappa^2=16\pi G$, and $\eta,\sigma,\xi$ take  the values +1 or -1. The quantities
$K$ and $K'$ are given by the following equations
\bea
K&=&R_{\mu\nu}R^{\mu\nu}-\frac{3}{8}R^2  \label{1}\\
K'&=&17R^3-72R_{\mu\nu}R^{\mu\nu}R+64R_{\mu}^{\nu}R_{\nu}^{\rho}R_{\rho}^{\mu}~.
\label{2}
\eea
\par\noindent
The equations of motion for $R^3$-NMG are written as \cite{Sinha:2010ai}
\be
\sigma G_{\mu\nu}+\Lambda
g_{\mu\nu}+\frac{1}{m^2}K_{\mu\nu}-\frac{\xi}{12\mu^4}K'_{\mu\nu}=0
\label{action}
\ee
\par\noindent
where
\bea
K_{\mu\nu}&=&g_{\mu\nu}(3R_{\alpha\beta}R^{\alpha\beta}-\frac{13}{8}R^2)+\frac{9}{2}RR_{\mu\nu}-8R_{\mu\alpha}R_{\nu}^{\alpha}+\frac{1}{2}(4\nabla^2R_{\mu\nu}-\nabla_{\mu}\nabla_{\nu}R-g_{\mu\nu}\nabla^2R)\\
\label{1}
K'_{\mu\nu}&=&17[-3R^2R_{\mu\nu}+3\nabla_{\mu}\nabla_{\nu}R^2+\frac{1}{2}g_{\mu\nu}R^3-3g_{\mu\nu}\nabla^2R^2]\nonumber\\
& &-72[-2RR_{\mu\alpha}R_{\nu}^{\alpha}-R_{\alpha\beta}R^{\alpha\beta}R_{\mu\nu}-\nabla^2(RR_{\mu\nu})+\nabla_{\mu}\nabla_{\nu}(R_{\alpha\beta}R^{\alpha\beta})
+2\nabla_{\alpha}\nabla_{(\mu}(R_{\nu)}^{\alpha}R)\nonumber\\
& &+\frac{1}{2}g_{\mu\nu}R_{\alpha\beta}R^{\alpha\beta}R-g_{\mu\nu}\nabla_{\alpha}\nabla_{\beta}(RR^{\alpha\beta})
-g_{\mu\nu}\nabla^2(R_{\alpha\beta}R^{\alpha\beta})]\nonumber\\
& &+64[-3R_{\mu}^{\rho}R_{\rho}^{\sigma}R_{\sigma\nu}-\frac{3}{2}\nabla^2(R_{\mu\alpha}R_{\nu}^{\alpha})+3\nabla_{\alpha}\nabla_{(\mu}(R_{\nu)}^{\beta}R_{\beta}^{\alpha})\nonumber\\
& &+\frac{1}{2}g_{\mu\nu}RR_{\alpha\beta}R^{\alpha\beta}-\frac{3}{2}g_{\mu\nu}\nabla_{\alpha}\nabla_{\beta}(R_{\rho}^{\alpha}R^{\rho\beta})]~.
\label{2}
\eea
It should be noted that in the limit $\mu\rightarrow \infty$, we recover the action and the corresponding equations
of motion for NMG \cite{Bergshoeff:2009hq}.
%
%
%
\section{$AdS_3$ vacuum in $R^3$-NMG}
%
%
%
\par\noindent
In this section, we will construct the $AdS_3$ vacuum solutions described by length scale $l^2$ in the context
of $R^3$-NMG. We assume  the global $AdS_3$ ansatz for the spacetime metric
\be
ds^2 =-\left(1+\frac{r^2}{l^2}\right)dt^2+\frac{dr^2}{\left(1+\frac{r^2}{l^2}\right)}+r^2d{\varphi}^2 ~.
\label{AdSmetric}
\ee
\par\noindent
By employing the coordinate redefinition $\varphi=\frac{x}{l}$ in Eq.  (\ref{AdSmetric}), the metric becomes
\be
ds^{2} = -\left(1+\frac{r^2}{l^2}\right)dt^2+\frac{dr^2}{\left(1+\frac{r^2}{l^2}\right)}+\frac{r^2}{l^2}dx^{2} ~.
\label{15}
\ee
\par\noindent
Substituting Eq. (\ref{AdSmetric}) in Eq. (\ref{action}), we obtain the equations of motion for the $AdS$ vacua
\be
4l^6m^2{\mu}^4{\Lambda}-260m^2{\xi}-l^2\left({\mu}^4+128m^2{\xi}\right)+4l^4m^2{\mu}^4{\sigma}=0 ~.
\label{AdSeom}
\ee
\par\noindent
It should be stressed that  apart from the fact that Eq. (\ref{AdSeom}) gives the form of the $AdS$ vacua, it also
provides  a constraint between the cosmological parameter, i.e. $\Lambda$, the $AdS$
length, i.e. $l$,  and the mass parameters, i.e. $m$ and $\mu$. This constraint is expressed through the formula
\be
\Lambda=-\frac{\sigma}{l^2}+\frac{1}{4{m^2}{l^4}}+\frac{\xi(65+32{l^2})}{{\mu^4}{l^6}}~.
\label{AdSLambda}
\ee
\par\noindent
For simplicity, we make the substitution  $y=l^2$ in Eq. (\ref{AdSeom}) and this equation transforms
into a third order polynomial equation. In this case, there are three $AdS$ vacua
\bea
l_1^2&=&-\frac{\sigma}{3\Lambda}+\frac{4B^{1/3}}{m^2{\mu}^4}+\frac{4E}{m^2{\mu}^4
B^{1/3}}=-\frac{\sigma}{3\Lambda}+\frac{4B^{2/3}+4E}{m^2{\mu}^4
B^{1/3}}\label{24}\\
l_2^2&=&-\frac{\sigma}{3\Lambda}-\frac{4D^{2/3}}{m^2{\mu}^4
B^{1/3}}-\frac{4E}{m^2{\mu}^4
B^{1/3}}+i\sqrt{3}\left[\frac{4D^{2/3}}{m^2{\mu}^4 B^{1/3}}-\frac{4
E}{m^2{\mu}^4 B^{1/3}}\right]\label{25}\\
l_3^2&=&-\frac{\sigma}{3\Lambda}-\frac{4D^{2/3}}{m^2{\mu}^4
B^{1/3}}-\frac{4 E}{m^2{\mu}^4
B^{1/3}}+i\sqrt{3}\left[-\frac{4D^{2/3}}{m^2{\mu}^4 B^{1/3}}+\frac{4
E}{m^2{\mu}^4 B^{1/3}}\right]\label{26}
\eea
where
\bea
B &=& -9m^4\Lambda{\mu}^{12}\sigma-4m^6{\mu}^8\left(-1755{\Lambda}^2\xi+288\Lambda\xi\sigma+2{\mu}^4{\sigma}^3\right) \nn\\
&&+\left\{-\left[3m^2{\Lambda}{\mu}^8+4m^4{\mu}^4\left(96{\Lambda}\xi+{\mu}^4{\sigma}^2\right)\right]^3 \right.\nn\\
&&+\left. m^8{\mu}^{16}\left[9{\Lambda}{\mu}^4\sigma+m^2\left(-7020{\Lambda}^2\xi
+1152\Lambda\xi{\sigma}+8{\mu}^4{\sigma}^3\right)\right]^2 \right\}^{1/2}~,\nn\\
D &=& -9m^4\Lambda{\mu}^{12}\sigma-4m^6{\mu}^8\left(-1755{\Lambda}^2\xi+288\Lambda\xi\sigma+2{\mu}^4{\sigma}^2\right)\nn\\
&&+3\sqrt{3}\left\{m^6{\Lambda}^2{\mu}^{12}\left[1825200m^6{\Lambda}^2{\mu}^4{\xi}^2-\Lambda{\left({\mu}^4+128m^2\xi\right)}^3 \right.\right.\nn\\
&&\left.\left.-4680m^4{\Lambda}{\mu}^4\left({\mu}^4+128m^2\xi\right)\sigma
+ m^2{\mu}^4{\sigma}^2\left(-{\left({\mu}^4+128m^2\xi\right)}^2-4160m^4{\mu}^4\xi\sigma\right)\right]\right\}^{1/2}~,\nn\\
E &=& 3m^2\Lambda{\mu}^8+4m^4{\mu}^4\left(96 \Lambda \xi +{\mu}^4{\sigma}^2\right)~.\nn
\eea
\par\noindent
Since we have obtained three maximally symmetric vacua, there should be, correspondingly,
 three effective cosmological constants, each one with the value $-l^{-2}_{i}$, where $i=1,2,3$.
However,  two of the vacua  have imaginary terms, namely Eqs. (\ref{25}) and (\ref{26}),
and, in general, squared quantities which obtain imaginary values are unacceptable.
It is known that for a third order equation real roots arise in the case of degeneracies and this is the reason
that, henceforth,  we are mainly concerned about  the existence of degeneracies.
\par\noindent
There is a triple degeneracy if and only if the cosmological parameter, $\Lambda$, satisfies simultaneously the relations
\bea
\Lambda_1&=&-\frac{4m^2{\mu}^4{\sigma}^2}{3\left({\mu}^4+128m^2\xi\right)}\label{Lambdadegen1}\\
\Lambda_2&=&\frac{9\sigma{\mu}^4+1152m^2\xi\sigma\pm\left[224640m^4{\mu}^4\xi{\sigma}^3
+\left(9\sigma{\mu}^4+1152m^2\xi\sigma\right)^2\right]^{1/2}}
{14040m^2\xi}~.
 \label{Lambdadegen2}
\eea
\noindent
The aforesaid condition, i.e. the equality between Eqs. (\ref{Lambdadegen1}) and (\ref{Lambdadegen2}), is true
if the following relation between the mass parameters holds
\be
{\mu}^4=-8\left(16m^2\xi+195m^4\xi\sigma\pm\sqrt{195m^6{\xi}^2\sigma\left(32+195m^2\sigma\right)}\right)~.
\label{massparamrelat}
\ee
\par\noindent
In addition, to avoid any imaginary values in Eq. (\ref{massparamrelat}),  we impose the condition
\be
m^2 \ge -\frac{32\sigma}{195 }~.
\label{msqeq}
\ee
\par\noindent
Moreover, from Eq. (\ref{Lambdadegen1}) the following constraint has to be satisfied
\be
\mu^4 \neq -128m^2\xi ~.
\ee
\par\noindent
Therefore, the  value of the triple degenerate vacuum is of the form
\be
l^2=-\frac{\sigma}{3\Lambda}~.
\ee
\par\noindent
At this point a couple of comments are in order.
First, in the framework of NMG \cite{Bergshoeff:2009hq,Giribet:2010ed}, the corresponding degenerate vacuum is of the form
\be
l^2_{NMG}=-\frac{\sigma}{2\Lambda}~.
\ee
\par\noindent
Thus, one can make the assumption that if higher order curvature terms of the form $R^{n}$ are
taken into account, then the unique degenerate vacuum will be of the form
\be
l^2=-\frac{\sigma}{n\Lambda}~.
\ee
\par\noindent
Second, it is obvious from  Eq. (\ref{massparamrelat}) that there are two triple degenerate vacua depending on the sign of
the last term of Eq. (\ref{massparamrelat}). Furthermore, in the case that the equality in Eq. (\ref{msqeq}) is fulfilled, there will be only
one triple degenerate vacuum.
\par\noindent
There is a double degeneracy if the cosmological parameter satisfies the condition
\be
\Lambda\neq-\frac{4m^2{\mu}^4{\sigma}^2}{3\left({\mu}^4+128m^2\xi\right)}
\ee
\par\noindent
and, additionally,
\bea
\Lambda &=& \frac{1}{3650400 m^6 \mu^4 } \left \{\mu^{12}+ 384 m^{2} \mu^{8} \xi + 49152 m^{4} \mu^{4} + 2097152 m^{6} \xi^{3}
+ 4680 m^{4} \mu^{8} \xi\sigma \right.\nonumber\\
& &\left. + 599040 m^{6} \mu^{4} \sigma
\pm \left [ \left ( \mu^{4}+ 128 m^{2} \xi \right )^{2} + 3120 m^{4} \mu^{4} \xi \sigma \right ]^{3/2} \right\}~.
\label{Lambdadoubledegen}
\eea
\par\noindent
In this case, the values of the $AdS_3$ vacua are
\bea
l^2_1&=&\frac{\mu^4\sigma+4m^2\xi\left(32\sigma-585\Lambda\right)}
{6\Lambda\left(\mu^4+128m^2\xi\right)+8m^2\mu^4} \label{ldoubledegen}\\
l^2_2&=&-\frac{\sigma}{\Lambda}+\frac{4m^2\xi\left(585\Lambda-32\sigma\right)-\sigma\mu^4}
{3\Lambda\left(\mu^4+128m^2\xi\right)+4m^2\mu^4} \label{lsimple}
\eea
\par\noindent
where Eq. (\ref{ldoubledegen}) corresponds to the double degenerate vacuum.
Substituting Eq. (\ref{Lambdadoubledegen}) into the Eqs. (\ref{ldoubledegen}) and (\ref{lsimple}),
the vacua read
\bea
l^2_1&=&\frac{\mu^4+128m^2\xi\pm \sqrt{\left(\mu^4+128m^2\xi\right)^2+3120m^4\mu^4\xi\sigma}}{4m^2\mu^4\sigma}\\
l^2_2&=&\frac{260m^2\xi\left[\mu^4+128m^2\xi\pm 2\sqrt{\left(\mu^4+128m^2\xi\right)^2+3120m^4\mu^4\xi\sigma}\right]}
{\left(\mu^4+128m^2\xi\right)^2+4160m^4\mu^4\xi\sigma} \label{l2squared}~.
\eea
\par\noindent
To avoid any imaginary values for the vacua,  certain bounds have to be imposed on the mass parameters. These bounds read
\bea
\frac{\mu^4+128m^2\xi}{m^2\mu^2} & \ge & \sqrt{3120\xi\sigma}\\
\frac{\mu^4+128m^2\xi}{m^2\mu^2} & \le & -\sqrt{3120\xi\sigma}
\eea
\par\noindent
where $\xi$ and $\sigma$ have to satisfy the condition $\xi\cdot \sigma=+1$.
Furthermore, in order to avoid any divergences and, thus, Eq. (\ref{l2squared}) to be well defined, the following constraint must hold
\be
\frac{\mu^4+128m^2\xi}{m^2\mu^2} \neq \pm \sqrt{4160\xi\sigma}~.
\ee
\par\noindent
In the context of Einstein Gravity, the asymptotic group of isometries on the boundary coincides with
the symmetries of two-dimensional conformal field theory (CFT) and, hence, the central
charge of the dual CFT can be computed \cite{Brown:1986nw}. In a similar way, the central charge
for parity preserving higher derivative gravity theories \cite{Nam:2010dd,Giribet:2010ed} reads
\be
c=\frac{l}{2G}g_{\mu\nu} \frac{\partial\mathcal{L}}{\partial
R_{\mu\nu}} \nonumber
\ee
\par\noindent
where $\mathcal{L}$ is the Lagrangian density in action given in Eq.  (\ref{action1}).
According to this expression, the central charge for $R^3$-NMG obtains the form \cite{AyonBeato:2009yq}
\be
c=\frac{3l}{2G} {\eta}\left[\sigma+\frac{1}{2m^2l^2}+\frac{\xi}{l^4{\mu}^4}\right]~.
\label{30}
\ee
\par\noindent
It is noteworthy that, in the parametric space, the point  where the central charge vanishes is treated as  a
special point for all higher derivative gravity theories in three dimensions. The reason is  that the  special point corresponds
to solutions in which logarithmic terms arise. These solutions are asymptotically $AdS_3$ solutions if one imposes relaxed boundary
conditions and, then, these solutions are dual to a Logarithmic CFT  (LCFT) \cite{Grumiller:2009sn}.
For the case of TMG and NMG, the special point is usually named chiral point \cite{Liu:2009kc,Henneaux:2009pw,Li:2008dq}.
Therefore, we are expecting an analogous behavior from  the corresponding special point in $R^3$-NMG.
It is evident from Eq. (\ref{30}) that the central charge  vanishes if the following relation is satisfied
\be
 \sigma+\frac{1}{2m^2l^2}+\frac{\xi}{l^4{\mu}^4}=0 ~.
\label{ccvanish}
\ee
\par\noindent
At this point, it should be stressed that we have obtained a very useful relation, i.e. Eq. (\ref{ccvanish}),
between the dimensionless parameters $m^2l^2$ and $\mu^4l^4$,
since it corresponds to one of the special points of $R^3$-NMG for which
the $AdS_3$ wave solutions will admit logarithmic terms, as will be shown in the next section.
%
%
%
\section{$AdS_3$ wave solutions in $R^3$-NMG}
%
%
%
\par\noindent
In this section, we derive exact gravitational wave solutions propagating
in $AdS_3$ space, in the context of $R^3$-NMG. Actually, the $AdS_3$ waves is
a special class of spacetimes defined when negative cosmological constant is present.
This class is called Siklos spacetimes \cite{siklos} and their special feature is the presence of a multiple
principal null-directed Killing vector,  $k^\mu$, of their Weyl tensor.
Here, we will follow the analysis of \cite{AyonBeato:2009yq,AyonBeato:2011qw}. Therefore,
we consider the metric of the $AdS_3$ waves
\be
ds^2=\frac{l^2}{y^2}\left[-F(u,y)du^2-2dud\upsilon+dy^2\right]~.
\label{AdSwaveansatz}
\ee
\par\noindent
This metric is conformally related to a pp-wave background \cite{Deser:2004wd, AyonBeato:2004fq}. The null
Killing vector reads $k^\mu \partial_\mu=\partial_\upsilon$. Here, the wave profile, i.e.
F, is an $\upsilon$-independent arbitrary function which determines the form
of the $AdS$ waves. An interesting fact here is that, if $F(u,y)=0$, we recover the
$AdS_3$ vacuum metric. On the other hand, if $F(u,y)\ll 1$, then Eq. (\ref{AdSwaveansatz})
refers to excitations around $AdS_3$ space. It should be noted  that in our analysis,
any linear, or quadratic, dependence of $F$ on the coordinate $y$ can be
eliminated due to a coordinate redefinition \cite{AyonBeato:2005qq} and, thus, we don't
take into account any such dependence.
\par\noindent
Inserting the $AdS_3$ wave ansatz, i.e. Eq. (\ref{AdSwaveansatz}), into Eq. (\ref{action}) and employing the expression
for the cosmological parameter as given in Eq. (\ref{AdSLambda}),  we obtain the following equation for the wave profile
\bea
\frac{\delta^u_\mu \delta^u_\nu}{4l^4m^2y^2\mu^4}
\left\{2\left(l^2\mu^4+4m^2\xi\right)\left(y^4\partial^4_yF+2y^3\partial^3_yF\right)\right.\nonumber\\
-\left.\left[6m^2\xi+l^2\mu^4\left(1-2m^2l^2\sigma\right)\right]\left(y^2\partial^2_yF-y\partial_yF\right)\right\}
&=& 0 ~.
\label{AdSwavediffequation}
\eea
\par\noindent
This is an Euler-Fuchs differential equation and it can be solved by making the substitution $F=y^a$.
In this case, we obtain a characteristic polynomial equation of the form
\be
a(a-2)\left[\left(a-1\right)^2-\frac{6m^2\xi+l^2\mu^4\left(1-2m^2l^2\sigma\right)}
{2\left(l^2\mu^4+4m^2\xi\right)}\right] = 0
\label{charpolynomial}
\ee
\par\noindent
and by defining the quantity
\be
M=\frac{6m^2\xi+l^2\mu^4\left(1-2l^2m^2\sigma\right)}{2\left(l^2\mu^4+4m^2\xi\right)}
\label{massgeneric}
\ee
\par\noindent
the characteristic polynomial equation, namely Eq. (\ref{charpolynomial}), is now written as
\be
a(a-2)\left[\left(a-1\right)^2-M\right] = 0
\label{charpolynomial2}
\ee
\par\noindent
Thus, the generic solution for the wave profile reads
\be
F(u,y)=F_+(u)y^{1+\sqrt{M}}+F_-(u)y^{1-\sqrt{M}}
\label{genericsolution}
\ee
\par\noindent
where $F_{\pm}(u)$ are arbitrary integration functions.
\par\noindent
As we have already mentioned in the previous section, there are special points
in the parametric space which correspond to solutions with logarithmic
terms and which are associated to a relaxed set of boundary
conditions. In the framework of TMG and NMG \cite{Giribet:2010ed,Grumiller:2009mw},
$AdS_3$ wave solutions with logarithmic wave profile have been obtained.
In order to derive a logarithmic wave profile, F, for the $AdS_3$ wave solutions in $R^{3}$-NMG,
the multiplicities of the characteristic polynomial equation, i.e. Eq. (\ref{charpolynomial2}), have to be obtained.
These multiplicities are computed at two special points which are defined through  relations that connect
the mass parameters of the theory.
\par\noindent
So, when $M=0$ in Eq. (\ref{massgeneric}), we get
\be
\sigma-\frac{1}{2m^2l^2}-\frac{3\xi}{\mu^4l^4}=0
\label{masslog1}
\ee
\par\noindent
and, from Eq. (\ref{charpolynomial2}), it is obvious that the root $a=1$ is of multiplicity 2. In this case,
the wave profile will be of the form
\be
F(u,y)=y\left[F_1(u)\ln{y}+F_2(u)\right]~.
\label{logprofile1}
\ee
\par\noindent
It should be noted that this logarithmic behavior is valid only if
\be
\mu^4l^4 \neq \frac{3\xi}{\sigma}~.\nn
\ee
\par\noindent
When $M=1$ in Eq. (\ref{massgeneric}), we get
\be
\sigma+\frac{1}{2m^2l^2}+\frac{\xi}{\mu^4l^4}=0
\label{masslog2}
\ee
and, from Eq. (\ref{charpolynomial2}), it is obvious that the roots $a=0$ and $a=2$ are of multiplicity 2.
In this case, both roots lead to the same wave profile of the form
\be
F(u,y)=\ln{y}\left[F_2(u)+F_4(u)y^2\right]~.
\label{logprofile2}
\ee
\par\noindent
It should be noted again that this logarithmic behavior is valid only if
\be
\mu^4l^4 \neq - \frac{\xi}{\sigma} \nn
\ee
\par\noindent
Finally, when $M<0$ in Eq. (\ref{massgeneric}), we obtain complex roots. In this case, we define the quantity
\be
M_1 =\frac{-6m^2\xi-l^2\mu^4\left(1-2m^2l^2\sigma\right)}{2\left(l^2\mu^4+4m^2\xi\right)}\nn
\label{Mcomplex}
\ee
and the generic solution for the wave profile, i.e. Eq. (\ref{genericsolution}), now becomes
\be
F(u,y)=y\left\{F_1(u)\cos\left(\sqrt{M_1}\ln{y}\right)
+F_2(u)\sin\left(\sqrt{M_1}\ln{y}\right)\right\}~.
\label{complexsol}
\ee
\par\noindent
The presence of the logarithmic terms leads to divergences from the asymptotically
$AdS_3$ spacetime which, consequently, leads to declinations from the conformal symmetry on the boundary.
In order to restore the appropriate asymptotic behavior for the $AdS_3$ wave solutions, namely its
Hamiltonian generators of the symmetries on the boundary to be represented by two
non-trivial copies of the infinite dimensional Virasoro algebra \cite{Brown:1986nw}, a relaxed set
of boundary conditions have to be imposed. Therefore, it is at the special points of the theory where
one has to impose the relaxed fall-off conditions and also a LCFT arises
as dual to the background of the $AdS_3$ waves in the bulk \cite{Grumiller:2009sn,Grumiller:2009mw}.
\par\noindent
It is evident from Eq. (\ref{genericsolution}) that the wave profile function satisfies the equation
\be
\Box F=m^2_{eff}F
\label{KleinGordon}
\ee
\par\noindent
where
\be
m^2_{eff}=M-\frac{1}{l^2}~.
\label{effectivemass}
\ee
\par\noindent
This is actually the Klein-Gordon equation and it means that the wave profile, i.e. $F$,
mimics a massive scalar field with effective mass $m^2_{eff}$ that propagates on the
$AdS_3$ wave background.
\par\noindent
Now, lets assume that $F_- = 0$, or $F_+=0$, in Eq. (\ref{genericsolution}).
Then, $F(u,y)$ will satisfy the Klein-Gordon equation which is now of the form, respectively,
\beas
\Box F(u,y)&=&m^2_+ F(u,y), \\
\mbox{or}\hspace{0.5cm} \Box F(u,y)&=&m^2_-F(u,y)
\eeas
\par\noindent
where
\be
m^2_+=m^2_-=M^2_{eff}=\frac{1}{l^2}\left(M-1\right)~.
\label{scalarmodesmass}
\ee
\par\noindent
We conclude that the generic solution given by Eq. (\ref{genericsolution}) expresses the
superposition of two massive scalar modes of the same effective mass propagating in the $AdS_3$ background.
\par\noindent
If we evaluate the aforesaid equations at the critical point given by Eq. (\ref{masslog1}),
then it is easily seen that the Klein-Gordon equation now reads
\be
\Box F=-\frac{1}{l^2}F
\label{scalarmodecriticalpoint1}
\ee
\par\noindent
and the effective mass will be
\be
m^2_{eff}=-\frac{1}{l^2}~.
\label{effmasscriticalpoint}
\ee
\par\noindent
It is obvious that the effective mass exactly saturates the Breitenlohner-Freedman
bound  for each one of the scalar modes \cite{Breitenlohner:1982jf}. Thus,
we are in complete accordance with the predictions in \cite{Setare:2013fza}
at the specific point. The same conclusion could be drawn by deriving
the Klein-Gordon equation from Eq.~(\ref{KleinGordon}) and imposing the condition $M=0$
in the expression for the effective mass given by Eq.~(\ref{effectivemass}).
\par\noindent
Concerning the second special point given by Eq. (\ref{masslog2}), the effective mass of the
two massive scalar modes vanishes since now $M=1$ (see Eq. (\ref{scalarmodesmass})).
This situation is completely different from the previous one since now the wave
profile given by Eq.~(\ref{logprofile2}) does not satisfy the Klein-Gordon equation anymore.
As a result, Eq.~(\ref{logprofile2}) can not be interpreted as a scalar field propagating
in the background under study.
\par\noindent
An interesting fact here is that the special point given by Eq. (\ref{masslog2})
can be identified as the point in the parametric space at which the central charge vanishes (see Eq. (\ref{ccvanish})).
Moreover, the same behavior is encountered at the special (chiral) point of TMG and NMG \cite{AyonBeato:2009yq,AyonBeato:2005qq,Liu:2009kc,AyonBeato:2004fq,Li:2008dq}.
\par\noindent
A couple of points are in order here. First, if the following two conditions are simultaneously satisfied
\bea
\mu^4l^4&=&-4 \xi m^2l^2 \\
\mu^4l^4&\neq&2m^2l^2 \left(\sigma\mu^4l^4-3\xi\right)
\eea
then from Eq.~(\ref{AdSwavediffequation}), the wave profile will now satisfy the equation
\be
\Box F=0
\ee
\par\noindent
with its solution to be
\be
F=C_0 y^2 ~.
\label{GRwavesolutions}
\ee
\par\noindent
As we have already pointed out, any quadratic dependence in three dimensions can be locally eliminated
due to a coordinate transformation. This means that the wave profile will take the form
$F=0$ , and, thus, the metric given in Eq. (\ref{AdSwaveansatz}) describes an $AdS_3$ spacetime.
In this case, there is no degree of freedom propagating in the background and the space is a trivial one \cite{Carlip:1998uc}.
\par\noindent
Second,  if the following two conditions are simultaneously satisfied
\bea
\mu^4l^4&=&-4 \xi m^2l^2 \\
\mu^4l^4&=&2m^2l^2 \left(\sigma\mu^4l^4-3\xi\right)
\eea
\par\noindent
which can be  merged into the following one condition
\be
\mu^4l^4=\frac{\xi}{\sigma}~,
\ee
then the differential equation given by Eq. (\ref{AdSwavediffequation}) is satisfied
for any wave profile function $F(u,y)$.
%
%
%
\section{Asymptotically Lifshitz black hole solutions}
%
%
%
\par\noindent
In this section, we attempt to derive black hole solutions which
are asymptotically Lifshitz in the context of $R^{3}$-NMG. This
derivation has already been performed for the case of NMG
\cite{AyonBeato:2009nh}. First, we  will follow an analogous analysis to
the one in Section III, in order to reveal the structure and
characteristics of the Lifshitz vacuum. For this reason, we utilize the
Lifshitz ansatz
\be
ds^2=-\frac{r^{2z}}{l^{2z}}dt^2+\frac{l^2}{r^2}dr^2+\frac{r^2}{l^2}dx^2\label{lifshitzansatz}
\ee
\par\noindent
and substitute it into Eq. (\ref{action}). Then, the equations of motion for the Lifshitz vacuum read
\footnotesize
\bea
\frac{1}{12}\frac{r^{2z}}{l^{2z}}\left\{-\frac{3\left(z^2-3z+1\right)
\left(z^2+z-1\right)}{l^4 m^2}-12\Lambda-\frac{12\sigma}{r^2}\right.\nonumber\\
\left.+\frac{4\xi\left[19+16l^2\left(1+z^2\right)\left(1+z+z^2\right)
+z\left(10z^5+30z^4+69z^3+4z^2+27z+36\right)\right]}{l^6{\mu}^4}\right\}&=&0     \label{Lifeom1}
\eea
\bea
\frac{1}{12r^2}\left\{\frac{3\left(z^2-3z+1\right)\left(z^2-z-1\right)}
{l^2 m^2}+12\Lambda l^2+12z\sigma\right.\nonumber\\
\left.+\frac{4\xi\left[10+16l^2\left(1+z^2\right)\left(1+z+z^2\right)
+z\left(10z^5+39z^4+66z^3-35z^2+66z+39\right)\right]}{l^4{\mu}^4}\right\}&=&0     \label{Lifeom2}\\
\frac{r^2}{12l^8}\left\{-\frac{3l^2\left(z^2-3z+1\right)\left(z^2-z-1\right)}
{m^2}+12\Lambda l^6+12z^2l^4\sigma\right.\nonumber\\
\left.-\frac{4\xi\left[10+16l^2\left(1+z^2\right)\left(1+z+z^2\right)
+z\left(19z^5+36z^4+27z^3+4z^2+69z+30\right)\right]}{{\mu}^4}\right\}&=&0        \label{Lifeom3}
\eea
\normalsize
\par\noindent
These equations of motion are satisfied for the following
values of the cosmological parameter $\Lambda$ and the mass parameter $m^2$
\bea
\Lambda&=&\frac{1}{6{\mu}^4l^6}\left\{\left[29+32l^2\left(1+z^2\right)
\left(z^2+z+1\right)\right.\right.\nonumber\\
&& \left.\left. +z\left(29z^5+75z^4+102z^3-25z^2+102z+75\right)\right]
-3{\mu}^4l^4\sigma\left(z^2+z+1\right)\right\} \label{Liflambdam}\\
m^2&=&\frac{{\mu}^4l^2\left(z^2-3z+1\right)}{2\sigma{\mu}^4l^4-2\xi
\left(z^2+z-3\right)\left(3z^2-z-1\right)} \label{Lifmsq}
\eea
\par\noindent
and can be rewritten as
\bea
\Lambda l^2&=&\frac{1}{6{\mu}^4l^4}\left\{\left[29+32l^2\left(1+z^2\right)
\left(z^2+z+1\right)\right.\right.\nonumber\\
&& \left.\left.+z\left(29z^5+75z^4+102z^3-25z^2+102z+75\right)\right]
-3{\mu}^4l^4\sigma\left(z^2+z+1\right)\right\} \label{dimensionlesslambda}\\
m^2 l^2&=&\frac{{\mu}^4l^4\left(z^2-3z+1\right)}{2\sigma{\mu}^4l^4-2\xi\left(z^2+z-3\right)
\left(3z^2-z-1\right)}
\label{dimensionlessmsq}
\eea
\par\noindent
where $\Lambda l^{2}$ and $m^{2} l^{2}$ are dimensionless quantities.
\par\noindent
If we evaluate Eqs. (\ref{dimensionlesslambda}) and (\ref{dimensionlessmsq})
at the limit $\mu\rightarrow\infty$ which is the limit where NMG is reinstated,
we exactly reproduce Eq. (7) of \cite{AyonBeato:2009nh}. Thus, we are in complete
accordance with the predictions of NMG at the limit where this theory shows up.
\par\noindent
Moreover, we determine the Lifshitz vacua through Eq. (\ref{Lifmsq}). In particular,
by solving Eq. (\ref{Lifmsq}), the Lifshitz length scale obtains two different
values
\footnotesize
\bea
l_-^2&=&\frac{1}{4m^2\sigma}\left\{\left(z^2-3z+1\right)-\sqrt{\left(z^2-3z+1\right)^2
+16 \frac{m^4}{\mu^4}\left(z^2+z-3\right)\left(3z^2-z-1\right)\xi\sigma}\right\}  \label{Lifvac1} \\
l_+^2&=&\frac{1}{4m^2\sigma}\left\{\left(z^2-3z+1\right)+\sqrt{\left(z^2-3z+1\right)^2
+16 \frac{m^4}{\mu^4}\left(z^2+z-3\right)\left(3z^2-z-1\right)\xi\sigma}\right\}\
\label{Lifvac2}
\eea
\normalsize
\par\noindent
which means that there are two Lifshitz vacua in the context of $R^3$-NMG.
On the other hand, there is a unique length scale in NMG \cite{Hohm:2010jc}. If we evaluate the
two vacua at the NMG limit, then the result reads
\bea
l^2_-&=&0 \label{LifvacNMGlim1} \\
l^2_+&=&\frac{z^2-3z+1}{2m^2 \sigma}~.
\label{LifvacNMGlim2}
\eea
\par\noindent
We observe that the length scale $l^2_+$ is identical with the one in NMG (see Eq. (5.58) in \cite{Hohm:2010jc}).
The zero value of the length scale $l^2_-$ indicates that the effective cosmological constant
$l^{-2}_{-}$ is infinite as well as the corresponding vacuum energy. Such infinities  are not
physically acceptable and, thus, the only scale that survives at the NMG limit is the length scale $l^2_{+}$ .
\noindent
Moreover, in order for the vacua not to admit imaginary terms, an additional condition has to be imposed
\be
\frac{16\xi\sigma m^4}{{\mu}^4} \ge -\frac{\left(z^2-3z+1\right)^2}
{\left(z^2+z-3\right)\left(3z^2-z-1\right)}~.
\label{imaginarybound}
\ee
\par\noindent
In the case where the two sides of Eq. (\ref{imaginarybound}) are equal, namely
\be
\frac{16\xi\sigma m^4}{{\mu}^4}=-\frac{\left(z^2-3z+1\right)^2}{\left(z^2+z-3\right)
\left(3z^2-z-1\right)}~,
\label{lifdegenvac}
\ee
the two Lifshitz vacua are degenerated
\be
l_-^2=l_+^2=l^2=\frac{\left(z^2-3z+1\right)}{4m^2\sigma}~.
\label{lifsinglevac}
\ee
\par\noindent
The latter equation provides the dimensionless quantity
\be
m^2l^2=\frac{\left(z^2-3z+1\right)}{4\sigma}~.
\label{m2l2}
\ee
\par\noindent
Substituting Eq. (\ref{m2l2}) into Eq. (\ref{lifdegenvac}), a second
dimensionless quantity is obtained
\be
\mu^4l^4=-\xi\sigma\left(z^2+z-3\right)\left(3z^2-z-1\right)~.
\label{mu4l4}
\ee
\par\noindent
Now, following the analysis in \cite{AyonBeato:2009nh}, we focus
on the special points of the theory in order to investigate whether there are, or not,
asymptotically Lifshitz black hole solutions in $R^{3}$-NMG, as has already been done in the context of NMG
\cite{AyonBeato:2009nh,Cai:2009ac,Gonzalez:2011nz,AyonBeato:2010tm}.
We adopt the following ansatz for the black hole metric
\be
ds^2=-\frac{r^{2z}}{l^{2z}}F(r)dt^2+\frac{l^2}{r^2}H(r)dr^2+\frac{r^2}{l^2}dx^2
\label{asymplifbh}
\ee
where the radial functions $F(r)$ and $H(r)$ are subjacent to the same
constraints that appear in \cite{AyonBeato:2009nh}, namely $F(r)=H(r)^{-1}$,
$ F(r)=H(r)^{-1}=1$ as $r \rightarrow \infty$, and there is an event horizon at $r_{H}$ that satisfies $F(r_{H})=H(r_{H})^{-1}=0$.
\par\noindent
The effective mass of the wave profile $F$ saturates the BF bound
at the critical point where Eq. (\ref{masslog1}) holds and now
takes the form
\be
m^2l^2=\frac{{\mu}^4l^4}{2\left(\sigma{\mu}^4l^4-3\xi\right)}~.
\label{criticalpoint2}
\ee
\par\noindent
Substituting Eq. (\ref{criticalpoint2}) into Eq. (\ref{dimensionlessmsq}), we get
\be
z\left(z-3\right)-\frac{\xi}{\sigma\mu^4l^4}\left[3\left(z^2-3z+1\right)
-\left(z^2+z-3\right)\left(3z^2-z-1\right)\right]=0 ~.
\label{equationz}
\ee
\par\noindent
This equation will give us the appropriate values of $z$ which are related to the
possible black hole solutions in $R^{3}$-NMG. An interesting feature of Eq. (\ref{equationz}) is that
if we take the NMG limit, i.e. $\mu\rightarrow \infty$, then Eq. (\ref{equationz}) becomes $z(z-3)=0$.
The root $z=3$ is the value of the dynamical exponent for the Lifshitz black hole in the NMG \cite{AyonBeato:2009nh}.
Furthermore, Eq. (\ref{equationz}) is a 4th order polynomial equation. The root $z=0$ corresponds to a generic
and trivial solution for any value of $\mu^4l^4$. Since Lifshitz vacuum solutions for imaginary values of the dynamical
exponent $z$ are not acceptable, we are interested in deriving the real roots of $z$.
\par\noindent
First, we investigate the case the three roots to be degenerated. However, this is true only if $\xi$ is zero.
But this fact contradicts to the definition of $\xi$ which can take only the value $+1$ or $-1$.
Moreover, if $\xi=0$ the theory reduces to NMG. Thus, there is no way to have a triple degenerated value of $z$.
Second, it is possible to have a double degenerated  value, i.e. $z_{1}$, and a simple root, i.e. $z_{2}$, respectively,
\bea
z_1&=&\frac{1}{18}\left(83+\frac{7865\xi}{9\sigma\mu^4l^4-130\xi}\right) \label{zdoubledegen}\\
z_2&=&-\frac{89}{9}+\frac{7865\xi}{1170\xi-81\sigma\mu^4l^4}~,
\label{zsingle}
\eea
\par\noindent
if and only if the mass parameter $\mu$ satisfies the following condition
\bea
\mu^4l^4&=&\frac{1}{8748}\left\{\frac{125519\xi}{\sigma}
-\frac{\left(9554589825301+1009110960\sqrt{92774435}\right)^{1/3}l^8\xi^2\sigma}
{\left(l^{24}\xi^3\sigma^6\right)^{1/3}}\right.\nonumber\\
&&\left.+\frac{\left(-9554589825301+1009110960\sqrt{92774435}\right)^{1/3}
\left(l^{24}\xi^3\sigma^6\right)^{1/3}}{l^8\sigma^3}\right\}~.
\eea
\par\noindent
Finally, we can obtain three simple roots among which only one is a real one.
For the case of the real simple root, the value of $z$ is
\be
z=-\frac{2}{9}+\frac{\left(260 \xi ^{2} - 18 l^{4} \mu ^{4} \xi  \sigma
+ 2^{1/3} A^{2/3}\right)}{9 \cdot 2^{2/3} \xi  A^{1/3}}
\label{zreal}
\ee
\par\noindent
where
\be
\hspace{-1.68ex} A \!=\! \left[-3445 \xi ^3+783 l^4 \mu ^4 \xi ^2 \sigma +
9 \sqrt{3} \sqrt{\xi ^3 \left(\xi -l^4 \mu ^4 \sigma \right)
\left(12675 \xi ^2-2015 l^4 \mu ^4 \xi  \sigma -12 l^8 \mu ^8 \sigma ^2\right)}\right]
\ee
\noindent
At this point, it is noteworthy that although the real value of $z$ depends only on the mass parameter $\mu$, it also fixes
the  value of the mass parameter, $m$, and that of the cosmological parameter, i.e. $\Lambda$.
\par\noindent
Finally,  if one substitutes one by one the specific values of $z$ as given in Eqs. (\ref{zdoubledegen},\ref{zsingle},\ref{zreal}),
into Eq. (\ref{asymplifbh})  and then introduces the obtained metrics into the equations
of motion as given by Eq. (\ref{action}), no solutions are obtained. This means that in the context of $R^{3}$-NMG there are not asymptotically
Lifshitz black holes solutions for the specific ansatz given by Eq. (\ref{asymplifbh}).
Furthermore, if one selects the form for the radial functions to be the one found in the context of NMG \cite{AyonBeato:2009nh}, i.e.
\be
F(r)=H(r)^{-1}= 1- \frac{M l^{2}}{r^2},
\ee
and introduces the obtained metric into the equations of motion in Eq. (\ref{action}), it is easily seen that there is no value of the
dynamical exponent z  for which the equations are fulfilled. This is in support to our result that there are not asymptotically
Lifshitz black holes solutions for the specific ansatz given by Eq. (\ref{asymplifbh}) in $R^{3}$-NMG.
%
%
%
\section{Conclusions}
%
%
%
\par\noindent
In this paper, we focus on three different but closely related aspects of  $R^{3}$-NMG. First, we discuss about the
$AdS_3$ vacua and the special points in the parametric space where degeneracies arises. Second, we obtain the
$AdS_3$ wave solutions which play a crucial role in our analysis of the $AdS_3$ structure, and finally
we search for asymptotically Lifshitz black hole solutions in $R^3$-NMG.
\par
In particular, we obtain the values of the $AdS_3$ vacua and search for the real degenerated values
since values with imaginary terms are not acceptable.  We show that imposing the appropriate conditions we can obtain a
triple degenerated vacuum as well as a double degenerated vacuum accompanied by a vacuum of multiplicity one. In addition,
we obtain the central charge of $R^{3}$-NMG.  The vanishing of the central charge is directly
related to the presence of logarithmic terms in $AdS$ wave solutions. For this reason, we derive and study the
$AdS_3$ wave solutions after substituting the $AdS$ wave ansatz into the equations of motion of $R^3$-NMG.
The generic solution depends from the wave profile function $F(u,y)$. The function F satisfies a Klein-Gordon equation and,
as a result,  it can be interpreted as a massive scalar field with effective mass $m_{eff}$ propagating in the $AdS_3$ space.
We proved that F is a superposition of two massive scalar modes with the same effective mass which both propagate in $AdS_3$ space.
The most significant result comes from the logarithmic branches that appear at specific points of the parametric space.
The first special point is defined through a characteristic polynomial equation when a quantity $M$ is equal to zero. In this case,
the effective mass saturates the Breitenlohner-Freedman bound and logarithmic terms appear in the $AdS_3$ wave solutions.
It is at this special point at which we have to impose the relaxed (weakened) boundary conditions so as to restore the appropriate asymptotic
behavior for the solutions. However, at this special point CFT is now a Logarithmic CFT which corresponds to the $AdS_3$ wave background.
The second special point is  also defined through the characteristic polynomial equation when the quantity $M$ is now equal to one.
It is this specific point at which the central charge vanishes. In addition, the generic wave profile can not anymore be interpreted
as a scalar field propagating in the $AdS_3$ wave background since it does not satisfy anymore the Klein-Gordon equation.
\par
Finally, we investigate  the existence of asymptotically Lifshitz black hole solutions in $R^{3}$-NMG by implementing the Lifshitz ansatz
into the corresponding equations of motion. We compute the values of the cosmological parameter, i.e. $\Lambda$, and the mass parameter,
i.e.  $m$, that satisfy the Lifshitz equations of motion. Concerning the dimensionless quantities $m^2l^2$ and
$\Lambda l^2$, we show that they exactly reproduce the corresponding formulae of NMG \cite{AyonBeato:2009nh} in the NMG limit.
We  obtain two Lifshitz vacua, namely $l^{2}_{+}$ and $l^{2}_{-}$. Taking the NMG limit, i.e. $\mu\rightarrow \infty$ we recover
the length scale of NMG, as expected. After evaluating the dimensionless quantity $m^2l^2$ at the first special point, we obtain a polynomial equation
which determines the values of the dynamical exponent $z$ as a function of the mass parameter $\mu$.
It is easily seen that,  by taking the NMG limit, the polynomial equation gives the value $z=3$ that determines the Lifshitz black hole solution in NMG.
We discover that there is no way to have a triple degenerate value of $z$. However, it is possible to have a double degenerate value of z
accompanied with one of multiplicity one, when  the mass parameter $\mu$ satisfies a specific condition. Moreover, it is possible to have
three simple roots but only one of them to be a real one. These values of the dynamical exponent $z$  correspond
to possible asymptotically Lifshitz black hole solutions. However, we show that for the specific values of the dynamical exponent $z$ and
a specific ansatz for the black hole metric, the equations of motion are not fulfilled. This proves that if one adopts the specific static ansatz for the black hole metric,
there are not any asymptotically Lifshitz black solutions in  $R^3$-NMG.
It is noteworthy that it may be possible one to obtain, with the same vales of the dynamical
exponent $z$, asymptotically Lifshitz black hole solutions in $R^3$-NMG by employing a specific non-static ansatz for the black hole metric.
We hope to address this issue in a future work.

%
%
%
%
%
%
\par\noindent
{\bf Acknowledgments}\\
ECV gratefully acknowledges discussion with P. Horava. MRS thanks G. Giribet for helpful discussions and correspondence.
GA and ECV thank T. Christodoulakis, T. Sotiriou, S. Nam, J-D Park, and S-H Yi  for useful correspondences.
%
%
%

\end{document}